\documentclass[12pt]{article}
\usepackage{amssymb,amsmath,epsfig}
\allowdisplaybreaks

\begin{document}
\title{\bf Dynamical Analysis of Self-gravitating Stars in Modified Gauss-Bonnet Gravity}

\author{ M. Z. Bhatti \thanks{mzaeem.math@pu.edu.pk}, Z. Yousaf \thanks{zeeshan.math@pu.edu.pk}, and A. Khadim \thanks{ammarakhadim4@gmail.com}\\
Department of Mathematics, University of the Punjab,\\
Quaid-i-Azam Campus, Lahore-54590, Pakistan.}

\date{}

\maketitle
\begin{abstract}
In this paper, we have continued the work of Herrera \emph{et al.}
\cite{herrera2012dynamical} in $f(G)$ gravitational theory. For this
purpose, a spherically symmetric fluid exhibiting locally
anisotropic pressure along with the energy density, is taken under
consideration. The perturbation scheme is imposed on modified field
equations and the dynamical equations. The collapse equation is
devised from these perturbed equations which stimulates to disclose
the instability zone under both Newtonian and post-Newtonian
constraints. It is wrapped up by concluding that dynamical
instability is interpreted by the adiabatic index $\Gamma$ which
relies on the anisotropic pressure, energy density and the dark
source terms due to $f(G)$ gravity.
\end{abstract}

\section{Introduction}

General relativity (GR) has become a vital constituent of the
present era which demonstrates the zigzagging of space time. This is
most alluring theory which explains all the gravitationally effected
phenomenon such as motion of the galaxy cluster, behavior of the
black holes and many such mysteries. General relativity lend a
different approach to visualize the universe and different
phenomenon happening in cosmos such as motion of the planets,
expansion of the universe and theoretical explanation of black
holes. Various experiments such as gravitational lensing,
gravitational redshift etc., were implemented to test GR and this
theory astonished by maintaining its standard. Despite of large
applications, however, there are some phenomenon which are not fully
illustrated by GR. Interior of black hole is one of such
limitations, since all physical laws are declined at singularity of
the black hole, as well as, GR has no quantum limit and unable to
clarify the dark matter and dark energy. So, collectively all these
shortcomings paved a path to develop of modified theories of gravity
which could help to explain all the queries related to dark energy
and the current cosmological model. Another incentive behind the
modified gravity is to consolidate different theories such as
Kaluza-Klein theory and string theory.

The modified Gauss-Bonnet gravity is one of the compelling modified
gravity theories. In modified Gauss-Bonnet gravity, $f(G)$ is the
general function of the Gauss-Bonnet with the Gauss-Bonnet invariant
$G$ and it helps to answer the queries related to late time cosmic
expansion. The $f(G)$ gravity theory illustrates dark energy with
much freedom as compared to other gravity theories. For suitable
choice of function $f$, this theory can justify the transition from
phantom to non-phantom state. The $f(G)$ gravity model was initially
proposed by Nojiri and Odintsov \cite{nojiri2005modified} by adding
some terms of Gauss-Bonnet function $f(G)$ to the Hilbert action
which helps to analyze various current features of cosmos. Baojiu
\emph{et al.} \cite{li2007cosmology} investigated the covariant and
gauge invariant perturbation equations and found that cosmological
data put few constraints on $f(G)$ models. Felice and Tsujikawa
\cite{de2009solar} studied the solar system constraints on
cosmologically feasible $f(G)$ gravity model and calculated some
corrections to the vacuum Schwarzschild solution. Also, they
performed some experiments for assessment of stability for the
modifications to GR. Goheer \emph{et al.}
\cite{goheer2009coexistence} presented the decelerating power-law
solutions for the particular form of $f(G)$ theories.

Garcia \emph{et al.} \cite{garcia2011f} dealt with a special model
$f(G)=\frac{a_1G^n+b_1}{a_2G^n+b_2}$ of Gauss-Bonnet gravity to
investigate the late time cosmic acceleration by imposing weak
energy conditions. Zhao \emph{et al.} \cite{zhao2012modified}
computed field equations and the equations of motion to figure out
non-conserved energy momentum tensor with effects of the specified
model, having the product of Lagrangian density and an arbitrary
function of Gauss-Bonnet term. Furthermore, they considered two
particular models $f(G)=\frac{a_1G^n+b_1}{a_2G^n+b_2}$ and
$f(G)=a_3G^n(1+b_3G^m)$ of $f(G)$ gravity to analyze the energy
conditions by the means of power law solution and equation of state
of matter with $\omega$ smaller than $-1/3$. Bamba \emph{et al.}
\cite{bamba2014bouncing} probed the bouncing cosmology and stability
conditions for its solution by remodeling $f(G)$ gravity. It was
also observed that unified model $F(G)=P(t)G+Q(t)$ helps to
scrutinize the late time cosmic acceleration along with the early
time bounce. Nojiri \emph{et al.} \cite{nojiri2010reconstruction}
constructed two action integrals, with and without auxiliary
scalars, to study cosmological reconstruction and
deceleration-acceleration transition in modified Gauss-Bonnet
gravity. It was found that this action integral corresponds to the
cosmological solutions having big bang and big rip singularity,
which contains auxiliary field.

The constraints under which $f(G)$ becomes cosmologically worthwhile
were extracted by Felice and Tusjikawa \cite{de2009construction}.
One of the imperative conditions for stability of late-time
de-Sitter solution is $d^2f/d^2G>0$ and it was found that
$d^2f/d^2G\rightarrow+0$ for $\mid G\mid\rightarrow\infty$ is
asymptotic behavior of feasible models. Zhou \emph{et al.}
\cite{zhou2009cosmological} made a study of $f(G)$ models and tested
several toy models to derive the conditions, for cosmologically
feasible $f(G)$ dark energy models, as geometrical constraints. As
well as useful trajectories, aping $\Lambda$CDM models in radiation
and matter dominated era, were also attained. Mohseni
\cite{mohseni2009non} studied the effects of force acting along the
four velocities of dynamic particles in the background of modified
Gauss-Bonnet gravity. Fayaz \emph{et al.} \cite{fayaz2015power}
examined the Gauss-Bonnet gravity in non-isotropic universe with the
help of power law solutions. They have explored the criteria for
transition to phantom phase and studied the stability issues of
modified gravitational models.

Rastkar \emph{et al.} \cite{houndjo2014exploring} proposed that when
the universe enters a phantom phase, a peculiar class of $f(G)$ has
power-law solutions irrespective of the matter dominated and
accelerating power-law solutions. Bamba \emph{et al.}
\cite{bamba2017energy} calculated viability conditions of some
particular $f(G)$ models induced by energy conditions with the help
of current estimated data of deceleration parameter. Bhatti \emph{et
al.} \cite{zaeem2018role} performed computational simulations to
check the stable regions of some strange stars with the help of
logarithmic $f(G,T)$ gravity, where $T$ indicates the trace of
matter tensor. Yousaf \cite{Yousaf2018,yousaf2019role} examined the
stability of collapsing stars and after evaluating structure
scalars, the role of Raychaudhuri equation is examined in this
context.

A physical model is worthwhile only if it is stable. The preeminent
eagerness of a star is to maintain its stability and so they undergo
the collapse or expansion due to struggle between the interstellar
pressure and gravitational pull in order to linger hydrostatic
equilibrium. Chan \emph{et al.} \cite{chan1993dynamical} studied the
influence of anisotropy and radiation on dynamical instability of
spherical system and found that stability of the system is highly
affected due to the influence of anisotropy and radiations. Chan
\emph{et al.} \cite{chan1994dynamical} explored the influence of
shear and shearing viscosity on dynamical instability of the
spherical fluid. They found that the fluid with shear collapses
rapidly where viscosity decreases the instability. Herrera \emph{et
al.} \cite{herrera1989dynamical} calculated the dynamical
instability ranges for the spherically symmetric non-adiabatic fluid
configuration in terms of $\Gamma$. Moreover, they formulated that
instability of the fluid increases or decreases due to the Newtonian
(N) corrections and the relativistic terms due to dissipation,
respectively.

Chan \emph{et al.} \cite{chan1989heat} calculated how the dynamical
instability is affected by the heat flow and found the range of
stability affected by the dissipation and relativistic corrections.
Alonso \emph{et al.} \cite{alonso1999heat} investigated the heat
conduction for the Lorentz gas by both the analytical and numerical
method and selected such crucial dynamical characters which are
related to the entire hyperbolic dynamics. Herminghaus
\cite{herminghaus1999dynamical} discussed that if the working
scenario of the two conducting plates is different then the
dielectric between them undergoes dynamical instability. Moreover,
he suggested that predicted output is result of the wave
intensifying. Herrera \emph{et al.} \cite{herrera2012dynamical}
performed stability analysis in order to calculate stability regimes
through perturbation scheme. Bhatti with his collaborators extended
their results and investigated the collapse rate for those
relativistic system that maintain plane
\cite{bhatti2017stability,doi:10.1142/S0217732319503334},
spherically symmetric
\cite{yousaf2017spherical,bhatti2019dissipative} and cylindrically
symmetric \cite{yousaf2016cavity,bhatti2017stabilitya} and axially
symmetric \cite{bhatti2020stability} geometries in their evolution
with modified gravity.

The formulation of the paper is as follows. In section \textbf{2},
the field equations and dynamical equations for spherically
symmetric anisotropic fluid in the background of $f(G)$ gravity are
explored. The perturbation scheme is introduced for metric and
material variables and applied to field equations in section
\textbf{3}. In section \textbf{4}, the perturbation is applied on
the dynamical equations and consequently the collapse equation is
calculated. In section \textbf{5}, the range of stability is
discussed at N and post-Newtonian (pN) eras.

\section{Fluid Distribution and Field Equations}

In order to continue a systematic investigation of dynamical
instability, we assume the fluid distribution to be spherically
symmetric and anisotropic. For such configuration, the line element
can be written as
\begin{equation}\label{1}
ds^2_-=-A^2dt^2+B^2dr^2+R^2(d\theta^2+\sin^2\theta d\phi^2),
\end{equation}
where $A(t,r),~B(t,r)$ and $R(t,r)$ are metric coefficients. The
energy-momentum tensor depicts the anisotropic fluid distribution of
the normal matter, mathematically given as
\begin{equation}\label{2}
T^-_{\alpha\beta}=(\mu+P_\bot)V_\alpha V_\beta +P_\bot g_{\alpha\beta}+(P_r+P_\bot)\chi_\alpha\chi_\beta,
\end{equation}
where $\mu$ is the energy density, $P_r$ the radial pressure,
$P_\bot$ the tangential pressure, $V^\alpha$ the four-velocity of
the fluid and $\chi_\alpha$ is unit four-vector along the radial
direction. The unit four-vectors satisfies the following identities
\begin{equation}\label{3}
V^\alpha V_\alpha=-1,\quad\chi^\alpha\chi_\alpha=1,\quad\chi^\alpha V_\alpha=0.
\end{equation}
We define the four-acceleration and the expansion of the fluid as
\begin{equation}\label{4}
a_\alpha=V_{\alpha;\beta} V^\beta,\quad\Theta=V^\alpha;_\alpha,
\end{equation}
where ; represents covariant derivative. The shearing motion in the
fluid can be characterized by the shear tensor whose mathematical
form is given by
\begin{equation}\label{5}
\sigma_{\alpha\beta}=V_{(\alpha;\beta)}+a_{(\alpha} V_{\beta)}-\frac{1}{3}\Theta(g_{\alpha\beta}+V_\alpha V_\beta).
\end{equation}
The four-vectors in comoving formalism satisfy
\begin{equation}\label{6}
V^a=A^{-1}\delta^a_0,\quad\chi^a=B^{-1}\delta^a_1.
\end{equation}
The non-zero components of four acceleration is evaluated to be
\begin{equation*}\label{7}
a_1=\frac{A'}{A},
\end{equation*}
where prime represents derivative with respect to $r$ while the
expression for expansion scalar found to be
\begin{equation}\label{8}
\Theta=\frac{1}{A}\left(\frac{\dot{B}}{B}+2\frac{\dot{R}}{R}\right)
\end{equation}
where dot represents the derivative w.r.t $t$. In a similar manner, the non-vanishing components of shear tensor are found as
\begin{equation}\label{9}
 \sigma_{11}=\frac{2}{3}B^2\sigma,\quad\sigma_{22}=\sigma_{33}\sin^{-2}\theta=-\frac{1}{3}R^2\sigma,
\end{equation}
while the shear scalar turns out to be
\begin{equation}\label{10}
\sigma=\frac{1}{A}\left(\frac{\dot{B}}{B}-\frac{\dot{R}}{R}\right).
\end{equation}
Now, we will take the modified Gauss-Bonnet gravity ($f(G)$) to
study its effects on the dynamical analysis of celestial objects.
This theory is viable because it passes the solar system tests under
some reasonable choice of $f(G)$ model and may describe the
late-time cosmic expansion \cite{nojiri2005modified}. This theory
leads to the discussion of phantom cosmic and quintessence
acceleration, cosmological constant with a chance to transit between
accelerated and decelerated phases in a quiet comprehensive way. The
Einstein-Hilbert action can be modified in this gravity theory as
\begin{equation}\label{11}
S=\int d^4x\sqrt{-g}\left(\frac{1}{2k}R+f(G)\right)+S_M (g_{\mu\nu},\psi),
\end{equation}
where $\kappa=8\pi$ and $S_M$ is the matter action indicating the
matter field depending upon the geometry while
$G=R^2-4R_{\mu\nu}R^{\mu\nu}+R_{\mu\nu\xi\sigma}R^{\mu\nu\xi\sigma}$
is the 4-dimensional topological invariant named as Gauss-Bonnet
invariant. Here, a specific function $f(G)=G+\alpha G^2$ of family
$f(G)$ is taken under consideration. Now, by varying above action
integral corresponding to metric tensor, we obtain following
modified field equations as
\begin{equation}\label{12}
G^-_{\alpha\beta}+D^-_{\alpha\beta}=T^-_{\alpha\beta},
\end{equation}
where
$G^-_{\alpha\beta}=R_{\alpha\beta}-{\frac{1}{2}}Rg_{\alpha\beta},$
is the Einstein tensor and
\begin{eqnarray}\nonumber
D^-_{\alpha\beta}&=&4R_{\alpha\rho\beta\sigma}\nabla^{\rho}\nabla^{\sigma}f_{G}+4(R_{\rho\beta}g_{\sigma\alpha}-R_{\rho\sigma}g_{\alpha\beta}
-R_{\alpha\beta}g_{\sigma\rho}+R_{\alpha\sigma}g_{\beta\rho})\nabla^{\rho}\nabla^{\sigma}f_{G}\\\label{13}&+&
2R(g_{\alpha\beta}g_{\sigma\rho}-g_{\alpha\sigma}g_{\beta\rho})\nabla^{\rho}
\nabla^{\sigma}f_G+{\frac{1}{2}}(Gf_{G}-f)g_{\alpha\beta},
\end{eqnarray}
with Ricci tensor $R_{\alpha\beta}$, Ricci scalar $R$, Riemannian
tensor $R_{\alpha\rho\beta\sigma}$ and the metric tensor
$g_{\alpha\beta}$. The non-trivial components of modified field
equations with Eqs.(\ref{1}),(\ref{2}),(\ref{6}) and (\ref{12}) are
given as
\begin{eqnarray}\nonumber
T^-_{00}&=&\mu{A}^2 =
\left(2\frac{\dot{B}}{B}+\frac{\dot{R}}{R}\right)\frac{\dot{R}}{R}-\left(\frac{A}{B}\right)^2\left[2\frac{R''}{R}
+\left(\frac{R'}{R}\right)^2-2\frac{B'}{B}\frac{R'}{R}-\left(\frac{B}{R}\right)^2\right]\\\label{14}&-&\frac{A^2}{2}
\left(Gf_G-f\right)+\chi_1\dot{f'}_G+\chi_2f''_G+\chi_3f'_G+\chi_4\dot{f}_G
,\\\label{15}
T^-_{01}&=&0=-2\left(\frac{\dot{R'}}{R}-\frac{\dot{B}}{B}\frac{\dot{R'}}{R}-\frac{\dot{R}}{R}\frac{A'}{A}\right)+\eta_1\dot{f}_G+\eta_2f'_G
+\eta_3\dot{f'}_G  ,\\\nonumber T^-_{11}&=&P_rB^2
=-\left(\frac{B}{A}\right)^2\left[2\frac{\ddot{R}}{R}-\left(2\frac{\dot{A}}{A}-\frac{\dot{R}}{R}\right)\frac{\dot{R}}{R}\right]+\left(2\frac{A'}{A}
+\frac{R'}{R}\right)\frac{R'}{R}\\\label{16}&-&\left(\frac{B}{R}\right)^2+\frac{B^2}{2}\left(Gf_G-f\right)+\phi_1\dot{f'}_G
+\phi_2{\ddot{f}}_G+ \phi_3\dot{f}_G+\phi_4f'_G  , \\\nonumber
T^-_{22}&=&T^-_{33}\sin^{-2}{\theta}
=P_{\bot}R^2=-\left(\frac{R}{A}\right)^2\left[\frac{\ddot{B}}{B}+\frac{\ddot{R}}{R}-\frac{\dot{A}}{A}\left(\frac{\dot{B}}{B}+\frac{\dot{R}}{R}\right)
+
\frac{\dot{B}}{B}\frac{\dot{R}}{R}\right]\\\nonumber&+&\left(\frac{R}{B}\right)^2\left[\frac{A''}{A}+\frac{R''}{R}-\frac{A'}{A}\left(\frac{B'}{B}-
\frac{R'}{R}\right)-\frac{B'}{B}\frac{R'}{R}\right]+\frac{R^2}{2}\left(Gf_G-f\right)\\\label{17}&+&\psi_1\ddot{f_G}+\psi_2{f''}_G
+\psi_3\dot{f}_G +\psi_4{f'}_G , \\\nonumber
\end{eqnarray}
where $\chi_i,~\eta_i,~\phi_i$ and $\psi_i$ are defined in the Appendix. The corresponding Ricci scalar is given as
\begin{eqnarray}\nonumber
R&=&-\frac{2}{R^2B^3A^3}(2R''BRA^3-2\ddot{R}B^3RA-\ddot{B}B^2R^2A+A''BR^2A^2\\\nonumber&-&\dot{R}^2B^3A-2\dot{R}\dot{B}RAB^2+2\dot{R}\dot{A}B^3R
+R'^2BA^3-2R'B'RA^3\\\label{18}&+&\dot{B}\dot{A}R^2B^2-B'A'R^2A^2-A^3B^3+2R'A'BRA^2).
\end{eqnarray}
The mass function $m(t,r)$ which describes the total energy inside
the radius $r$ is provided by
Misner-Sharp\cite{misner1964relativistic}. For our line element, it
turns our to be
\begin{equation}\label{19}
m=\frac{R}{2}\left[\left(\frac{\dot{R}}{A}\right)^2-\left(\frac{R'}{B}\right)^2+1\right].
\end{equation}
The non-trivial components of Bianchi identities using Eq.(\ref{12}) turns out to be
\begin{eqnarray}\nonumber
(T^{-\alpha\beta}_{;\beta}-D^{-\alpha\beta}_{;\beta})V_\alpha
&=&-\frac{1}{A}\left[\dot{\mu}+(\mu+P_r)\frac{\dot{B}}{B}+2(\mu+P_\bot)\frac{\dot{R}}{R}\right]-\frac{1}{A}\left[(-2\frac{\dot{A}}{A^3}
\right.\\\nonumber&+&\left.\frac{\dot{B}}{{A^2}B}+2\frac{\dot{R}}{R{A^2}})[\frac{A^2}{2}(Gf_G-f)-\chi_1{\dot{f'}}-\chi_2{f''}_G-\chi_3
{f'}_G\right.\\\nonumber&-&\left.\chi_4{\dot{f}_G}]+\frac{1}{A^2}[A\dot{A}(Gf_G-f)+\frac{A^2}{2}(\dot{G}f_G+G\dot{f}_G-\dot{f})
\right.\\\nonumber&-&\left.\dot{\chi_1}\dot{f'}_G-\chi_1\ddot{f'}_G-\dot{\chi}_2f''_G
-\chi_2\dot{f''}_G-\dot{\chi}_3{f'}_G-\chi_3\dot{f'}_G\right.\\\nonumber&-&\left.\dot{\chi}_4\dot{f}_G-\chi_4\ddot{f}_G]
+(\frac{B'}{B^3}-2\frac{R'}{{B^2}R})[-\eta_1\dot{f}_G-\eta_2{f'}_G-\eta_3\dot{f'}_G]\right.\\\nonumber&-&\left.\frac{1}{B^2}[-{\eta'}_1\dot{f}_G-
\eta_1\dot{f'}_G-\eta'_2{f'}_G-
\eta_2{f''}_G-{\eta'}_3\dot{f'}_G-\eta_3\dot{f''}_G]\right.\\\nonumber&+&\left.\frac{\dot{B}}{B^3}[-\frac{B^2}{2}(Gf_G-\dot{f})
-\phi_1\dot{f'}_G-\phi_2\ddot{f}_G-\phi_3\dot{f}_G-\phi_4f'_G]\right.\\\nonumber&+&\left.\frac{2\dot{R}}{R^3}[-\frac{R^2}{2}(Gf_G-f)
-\psi_1\ddot{f}_G-\psi_2{f''}_G-\psi_3\dot{f}_G-\psi_4{f'}_G]\right]\\\label{20}&=&0,
\\\nonumber
(T^{-\alpha\beta}_{;\beta}-D^{-\alpha\beta}_{;\beta})\chi_\alpha
&=&\frac{1}{B}\left[P'_r+(\mu+P_r)\frac{A'}{A}+2(P_r-P_\bot)\frac{R'}{R}\right]+\frac{1}{B}\left[\frac{1}{A^2}[\dot{\eta}_1\dot{f}_G
\right.\\\nonumber&+&\left.\eta_1\ddot{f}_G
+\dot{\eta}_2{f'}_G+\eta_2\dot{f'}_G+\dot{\eta}_3\dot{f'}_G+\eta_3\ddot{f'}_G]+\left(\frac{\dot{A}}{A^3}\right.\right.\\\nonumber&-&\left.\left.
2\frac{\dot{R}}{R{A^2}}\right) [-\eta_1\dot{f}_G-\eta_2{f'}_G-
\eta_3\dot{f'}_G]+(-\frac{2B'}{B^3}+\frac{A'}{A{B^2}}\right.\\\nonumber&+&\left.\frac{2R'}{R{B^2}})\left(-\frac{B^2}{2}(Gf_G-f)
-\phi_1\dot{f'}_G-\phi_2\ddot{f}_G-\phi_3\dot{f}_G\right.\right.\\\nonumber&-&\left.\left.
\phi_4{f'}_G\right)+\frac{1}{B^2}\left(-BB'(Gf_G-f)-\frac{B^2}{2}({G'}f_G-G{f'}_G\right.\right.\\\nonumber&-&\left.\left.{f'}
)-{\phi'}_1\dot{f'}_G-\phi_1\dot{f''}_G-{\phi'}_2\ddot{f}_{G}
-\phi_2\ddot{f'}_G-\phi_3\dot{f'}_G-{\phi'}_3\dot{f}_G\right.\right.\\\nonumber&-&\left.\left.\phi_4{f''}_G\right)+\frac{A'}{A^3}\left(\frac{A^2}{2}
(Gf_G-f)-\chi_1\dot{f'}_G-\chi_2{f''}
_G-\chi_3{f'}_G\right.\right.\\\nonumber&-&\left.\left.\chi_4\dot{f}_G\right)+2\frac{R'}{R^3}\left(\frac{R^2}{2}(Gf_G-G)+\psi_1\ddot{f}_G
+\psi_2{f''}_G+\psi_3\dot{f}_G\right.\right.\\\label{21}&+&\left.\left.\psi_4{f'}_G\right)\right]=0.
\end{eqnarray}
The extra curvature ingredients appearing in the above equations are given in the Appendix.

\section{The Perturbation Scheme }

In this section, we will apply perturbation on the field equations
and the dynamical equations which is the small change in the
physical system. Since, we are interested in analyzing the stability
or instability of the system after slightly disturbing the system by
means of perturbation. So, initially, we consider that fluid and
geometry is described only by radial dependent coordinates, i.e.,
the system is in static equilibrium. We also assume that with the
passage of time all the quantities have both radial and time
dependence. Also, we consider that $0<\epsilon\ll1$.

We are dealing with the non-linear partial differential equations.
In order to have deep analysis to see the role of radially dependent
variables on the dynamical instability of relativistic matter
content, we use a particular mathematical method to solve the
equations. In literature, there are very few methods of solving
non-linear differential equations, among them there is a method of
transforming subsequent equation into separable form
\cite{odibat2008differential,sharif2015role}. The mathematical
profile of the $f(G)$ model after a time $t=0$ can be expressed in a
separable form. Therefore, we have
\begin{eqnarray}\label{22}
A(t,r)&=& A_0{r}+\epsilon T(t)a(r),\\\label{23} B(t,r) &=&
B_0(r)+\epsilon T(t)b(r),\\\label{24}
R(t,r)&=&R_0(r)+{\epsilon}T(t)c(r),\\\label{25}
\mu(t,r)&=&\mu_0(r)+{\epsilon}\bar{\mu}(t,r),\\\label{26}
P_r(t,r)&=&P{r0}(r)+{\epsilon}\bar{P_r}(t,r),\\\label{27}
P_\bot(t,r)&=&P_{\bot0}(r)+{\epsilon}\bar{P_\bot}(t,r),\\\label{28}
m(t,r)&=&m_0(r)+{\epsilon}\bar{m}(t,r),\\\label{29}
\Theta(t,r)&=&{\epsilon}\bar{\Theta}(t,r),\\\label{30}
\sigma(t,r)&=&{\epsilon}\bar{\sigma}(t,r),\\\label{31}
G(t,r)&=&G_0(r)+{\epsilon}T(t)g(r),\\\label{32}
f(t,r)&=&G_0(1+{\alpha}G_0)+{\epsilon}Tg(1+2{\alpha}G_0).
\end{eqnarray}
We assume that the spherically symmetric anisotropic system with an
environment of $f(G)$ gravity is in a static state at a very large
past time that can be described through an equation $T(-\infty)=0$,
thereby putting $f(G)$ as a radial dependent function. After this,
the system equipped with $f(G)$ dynamics enters in the present state
with the passage of time and continues to collapse and moves on by
decreasing its areal radius.

Where the quantities with subscript zero only depend on $r$.
Equations (\ref{14})-(\ref{17}) by using Eqs.(\ref{22})-(\ref{32})
for static configuration turns out to be
\begin{eqnarray}\nonumber
{\mu}_0&=&\frac{1}{(r{B_0})^2}\left(2r\frac{{B'}_0}{B_0}+{B_0}^2-1\right)-\frac{\alpha{G_0}^2}{2}+
\frac{8\alpha{G_0''}}{{r^2}{B_0}^4}\left({1-{B_0}^2}\right)\\\label{33}&+&\frac{8{B'_0}\alpha{G'_0}}{{r^2}{B_0}^5}\left({{B_0}^2-3}\right),\\\nonumber
P_{r0}&=&\frac{1}{(rB_0)^2}\left(2r\frac{A'_0}{A_0}-{B_0}^2+1\right)+\frac{\alpha{G_0}^2}{2}+
\frac{8\alpha{G'_0}}{{r^2}{B_0}^4}\left(-3\frac{A'_0}{A_0}\right.\\\label{34}&+&\left.\frac{{B_0}^2{A'_0}}{A_0}\right),\\\nonumber
P_{\bot0}&=&\frac{1}{{B_0}^2}\left(\frac{A''_0}{A_0}-\frac{B'_0}{rB_0}
-\frac{{A'_0}{B'_0}}{A_0B_0}+\frac{A'_0}{A_0r}\right)+\frac{\alpha{G_0}^2}{2}
-\frac{8\alpha{A'_0}{G''_0}}{rA_0{B_0}^4}\\\label{35}&+&\frac{8}{rA_0{B_0}^4}
\left(3\alpha{B'_0}{A'_0}{G'_0}-{\alpha}{A''_0}B_0{G'_0}\right),
\end{eqnarray}
and for perturbed configuration, from Eq.(\ref{14})-(\ref{17}), we have
\begin{eqnarray}\nonumber
\bar{\mu}&=&-2\frac{T}{{B_0}^2}\left[\left(\frac{c}{r}\right)''-\frac{1}{r}\left(\frac{b}{B_0}\right)'-\left(\frac{B'_0}{B_0}-
\frac{3}{r}\right)\left(\frac{c}{r}\right)'
-\left(\frac{B_0}{r}\right)^2\left(\frac{b}{B_0}\right.\right.\\\nonumber&-&\left.\left.\frac{c}{r}\right)\right]-2{\mu_0}T\frac{b}{B_0}
-T\alpha\left[gG_0+\frac{b{G_0}^2}{B_0}
-\frac{4}{{r^2}{B_0}^4} \left(4r{G''_0}(\frac{c}{r})-\frac{b{G''_0}}{2B_0}\right.\right.\\\nonumber&+&\left.\left.4c\frac{{B_0}^2{G''_0}}{r}+2{g''}(1
-{B_0}^2)\right)-4\frac{B'_0}{{r^2}{B_0}^5}
\left(-12r{G'_0}(\frac{c}{r})'+18\frac{b{G'_0}}{B_0}\right.\right.\\\nonumber&-&\left.\left.2{g'}\left(3-{B_0}^2\right)
+6\frac{{b'}{G'_0}}{B'_0}-4c\frac{{G'_0}{B_0}^2}{r}-8bB_0{G'_0}
+2\frac{{b'}{G'_0}{B_0}^2}{B'_0}\right.\right.\\\label{36}&+&\left.\left.4{c''}{G'_0}{B_0}^3\right)\right],
\\\nonumber
0&=&2\frac{\dot{T}}{A_0B_0}\left[\left(\frac{c}{r}\right)'-\frac{b}{rB_0}-\left({\frac{A'_0}{A_0}-\frac{1}{r}}\right)\frac{c}{r}\right]+
8{\alpha}g\frac{\dot{T}}{{A_0}^4{B_0}^4}\left[A_0B_0{A'_0}{A''_0}\right.\\\nonumber&-&\left.\frac{{A_0}^2B_0{A'_0}}{r^2}A_0{A'_0}^2{B'_0}+
\frac{{A_0}^2{A'_0}{B_0}^3}{r^2}\right]+8\alpha\frac{\dot{T}{G'_0}}{r^2A_0^4B_0^4}\left[-A_0^3B_0c'+A_0^2{A'_0}B_0c\right.\\\nonumber&+&\left.
br^2A_0^2{A''_0}
-r^2A_0^2{A'_0}\frac{b{B'_0}}{B_0}+bA_0^3B_0^2\right]+8\alpha{g'}\frac{\dot{T}}{r^2A_0^3B_0^3}
\left[a_0^2+r^2A_0{A'_0}\frac{B'_0}{B_0}
\right.\\\label{37}&-&\left.A_0^2B_0^2+r^2A_0{A''_0}\right], \\\nonumber
\bar{P_r} &=&-2\frac{\ddot{T}}{{A_0}^2}\frac{c}{r}+2\frac{T}{r{B_0}^2}\left[\left(\frac{c}{r}\right)'\left(1+r\frac{A'_0}{A_0}\right)+\left(\frac{a}{A_0}\right)'
-\frac{B_0^2}{r}\left(\frac{b}{B_0}-\frac{c}{r}\right)\right]\\\nonumber&-&2P_{r0}T\frac{b}{B_0}+T\alpha\left[gG_0+b\frac{{G_0}^2}{B_0}\right]+
8\alpha\frac{g\ddot{T}}{r^2A_0^3B_0^3}\left[2r^2{A''_0}B_0-2r^2{A'_0}{B'_0}\right.\\\nonumber&+&\left.A_0B_0-4rA_0{B'_0}-A_0B_0^3\right]
+4\alpha\frac{{A'_0}T}
{r^2B_0^4A_0}\left[-12r{G'_0}\left(\frac{c}{r}\right)'+12b\frac{G'_0}{B_0}-6{g'}\right.\\\nonumber&-&\left.6{G'_0}\left(\frac{A_0}{A'_0}
\right)\left(\frac{a}{A_0}\right)'
-4{G'_0}B_0^2\frac{c}{r}+2{G'_0}B_0^2\left(\frac{A_0}{A'_0}\right)\left(\frac{a}{A_0}\right)'+2{g'}B_0^2\right]
\\\label{38}&+&16\alpha\frac{{G'_0}\ddot{T}}{A_0^2B_0^2}\frac{c}{r^2} ,  \\\nonumber
\bar{P_\bot}&=&-\frac{\ddot{T}}{A_0^2}\left(\frac{c}{r}+\frac{b}{B_0}\right)+\frac{T}{B_0^2}\left[\left(\frac{a}{A_0}\right)''+\left(\frac{c}{r}\right)''
+\left(2\frac{A'_0}{A_0}-\frac{B'_0}{B_0}+\frac{1}{r}\right)\left(\frac{a}{A'_0}\right)'\right.\\\nonumber&-&\left.\left(\frac{A'_0}{A_0}+
\frac{1}{r}\right)\left(\frac{b}{B_0}\right)'+\left(\frac{A'_0}{A_0}-\frac{B'_0}{B_0}+\frac{2}{r}\right)\left(\frac{c}{r}\right)\right]
-2P_{\bot0}T\frac{b}{B_0}+T\alpha(gG_0\\\nonumber&+&b\frac{G_0^2}{B_0})-8\alpha\frac{\ddot{T}}{rA_0^2B_0^3}(g{B'_0}-cB_0{G''_0}+c{B'_0}-b)-
8\alpha\frac{T{A'_0}}{rA_0B_0^4}\left[{c'}{G''_0}+g''\right.\\\nonumber&+&\left.{G''_0}\frac{A_0}{A'_0}(\frac{a}{A_0})'\right]+2T\alpha{b}
\frac{A'_0G''_0}{rA_0B_0^5}+8\alpha\frac{T}{rA_0B_0^5}\left[{A'_0}B_0{c''}{G'_0}+3{A'_0}{B'_0}\left({g'}\right.\right.\\\nonumber&-&\left.
\left.3b\frac{G'_0}{B_0}+
{c'}{G'_0}+{b'}\frac{G'_0}{B'_0}+{G'_0}\frac{A_0}{A'_0}\left(\frac{a}{A_0}\right)\right)-{A''_0}B_0\left({g'}+a\frac{G'_0}{A_0}-3b\frac{G'_0}{A_0}\right.
\right.\\\label{39}&+&\left.\left.{a''}
\frac{G'_0}{A_0}+b\frac{G'_0}{B_0}+{c'}{G'_0}\right)\right].\\\nonumber
\end{eqnarray}
The perturbation of expansion scalar from Eq.(\ref{8}) and shear from Eq.(\ref{10}) yields
\begin{eqnarray}\label{40}
\bar{\Theta}&=&\frac{\dot{T}}{A_0}\left(\frac{b}{B_0}+2\frac{c}{r}\right),\\\label{41}
\bar{\sigma}&=&\frac{\dot{T}}{A_0}\left(\frac{b}{B_0}-\frac{c}{r}\right),
\end{eqnarray}
while the static parts vanishes.

\section{The Collapse Equation}

The collapse equation will help to find the stability ranges for the
anisotropic fluid. By making use of perturbation on the Bianchi
identities given in Eqs.(\ref{20}) and (\ref{21}) with
Eqs.(\ref{22})-(\ref{32}), we obtain
\begin{eqnarray}\nonumber
0&=& \frac{1}{B_0}\left[{P'_{r0}}+(\mu_0+P_{r0})\frac{A'_0}{A_0}+\frac{2}{r}(P_{r0}-P_{\bot0})\right]+\left(\frac{A'_0}{A_0B_0^3}
-2\frac{B'_0}{B_0^4}+\frac{2}{rB_0^3}\right)
\left[\right.\\\nonumber&-&\left.\frac{B_0^2}{2}
\alpha
G_0^2-8\alpha\frac{G'_0}{r^2A_0B_0^2}(-3{A'_0}+B_0^2{A'_0})\right]+\frac{A'_0}{A_0^3B_0}\left[\alpha{G_0^2}\frac{A_0^2}{2}
-8\alpha\frac{G''_0} {r^2B_0^4}\right.\\\nonumber&\times&\left.(A_0^2-A_0^2B_0^2)-8\alpha\frac{G'_0}{r^2A_0B_0^5}(-3{B'_0}A_0^3+A_0^3B_0^2{B'_0})\right] +\frac{1}{B_0^3}
(-\alpha B_0{B'_0}{G_0^2}\\\nonumber&-&\alpha
B_0^2G_0{G'_0})-8\alpha\frac{G'_0}{(r^2A_0^3B_0^2)^2}\left[r^2\frac{A_0^3}{B_0}(-3{A_0^2}{A''_0}-6{A_0}{A'_0}^2+2{A_0}^2{B_0}
{A'_0}{B'_0}\right.\\\nonumber&+&\left.2A_0{A'_0}^2{B_0^2}+A_0^2{A''_0}B_0^2)-\left(2r\frac{A_0^3}{B_0}
+2r^2A_0^3\frac{B'_0}{B_0^2}+3r^2{A'_0}
\frac{A_0^2}{B_0}\right)(-3A_0^2{A'_0}\right.\\\nonumber&+&\left.A_0^2B_0^2{A'_0})\right]
-8\alpha\frac{G''_0}{r^2A_0^3B_0^5}(-3A_0^2A'_0+A_0^2A'_0B_0^2)+\frac{2}{r^3B_0}\left[r^2\alpha\frac{G_0^2}{2}
\right.\\\label{42}&-&\left.8rA_0^2{A_0'}\alpha\frac{G''_0}{A_0^3B_0^4}
+8\alpha\frac{G'_0}{A_0^3B_0^5}(3rA_0^2A'_0B'_0-rA_0^2B_0A''_0)\right],
\\\label{43}
0 &=&\frac{1}{A_0}\left[\dot{\bar{\mu}}+(\mu_0+P_{r0})\dot{T}\frac{b}{B_0}+2(\mu_0+P_{\bot0})\dot{T}\frac{c}{r}+D_1\dot{T}\right],\\\nonumber
0&=&\frac{1}{B_0}\left[\bar{P'_r}+(\mu_0+P_{r0})T(\frac{a}{A_0})'+(\bar{\mu}+\bar{P_r})\frac{A'_0}{A_0}+2(P_{r0}\right.\\\label{44}
&-&\left.P_{\bot0})T(\frac{c}{r})'
+\frac{2}{r}
(P_r-P_\bot)\right]+ D_2\ddot{T}+D_3T,
\end{eqnarray}
where $D_1,~D_2$ and $D_3$ are the dark source terms, appearing due
to $f(G)$ gravity, are defined in Appendix. Integration of
Eq.(\ref{43}) w.r.t $t$ gives
\begin{equation}\label{45}
\bar{\mu}=-\left[(\mu_0+P_{r0})\frac{b}{B_0}+2(\mu_0+P_{\bot0})\frac{c}{r}+D_1\right]T.
\end{equation}
The linear perturbation on the mass function $m(t,r)$ from Eq.(\ref{19}) leads to
\begin{eqnarray}\label{46}
m_0 &=&\frac{r}{2}\left(1-\frac{1}{B_0^2}\right),\\\label{47}
\bar{m}&=&-\frac{T}{B_0^2}\left[r\left(\bar{c}-\frac{b}{B'_0}\right)+(1-B_0^2)\frac{c}{2}\right].
\end{eqnarray}
The perturbed configuration of Ricci scalar given in Eq.(\ref{18}) gives a second order differential equation as follows
\begin{equation}\label{48}
\ddot{T}-\omega T=0,
\end{equation}
here
\begin{eqnarray}\nonumber
\omega&=&\left(2\frac{c}{rA_0^2}+\frac{b}{A_0^2B_0}\right)^{-1}\left[2\frac{c''}{rB_0^2}+\frac{a''}{A_0B_0^2}-2\frac{A''_0}{A_0}
\frac{b}{B_0^3}-a\frac{A''_0}{A_0^2B_0^2}+2\frac{c'}{r^2B_0^2}\right.\\\nonumber&-&\left.2\frac{b}{r^2B_0^3}-2\frac{c}{r^3B_0^2}
-2\frac{c'}{r}\frac{B'_0}{B_0^3}
-2\frac{b'}{rB_0^3}+2\frac{c}{r^2}\frac{B'_0}{B_0^3}+6\frac{bB'_0}{r^2B_0^4}-\frac{b'}{b_0^3}\frac{A'_0}{A_0}\right.\\\nonumber&-&\left.
\frac{a'}{A_0}\frac{B'_0}
{B_0^3}+a\frac{A'_0}{A_0^2}\frac{B'_0}{B_0^3}+3b\frac{A'_0}{A_0}\frac{B'_0}{B_0^4}+2\frac{c}{r^3}+
2\frac{A'_0c'}{rA_0B_0^2}+2\frac{a'}
{rA_0B_0^2}-2\frac{bA'_0}{rA_0B_0^3}\right.\\\label{49}&-&\left.2\frac{cA'_0}{r^2A_0B_0^2}-2\frac{aA'_0}{rA_0^2B_0^2}\right].
\end{eqnarray}
The solution of the above equation contains oscillating and
non-oscillating parts corresponding to stable and unstable
configurations of the stellar interior. Since, we are interested to
determine the dynamical instability so we neglect one part so that
the solution takes the form as
\begin{equation}\label{50}
T(t)=-\exp\left(\sqrt{\omega} t\right),
\end{equation}
which corresponds to the unstable (non-oscillating) function. We
consider second law of thermodynamics which relates $\bar{P}$ and
$\bar{\mu}$ via adiabatic index as
\begin{equation}\label{51}
\bar{P_r}=\Gamma\frac{P_{r0}}{\mu_0+P_{r0}}\bar{\mu},
\quad \bar{P_\bot}=\Gamma\frac{P_{\bot 0}}{\mu_0+P_{\bot 0}}\bar{\mu},
\end{equation}
where the adiabatic index, also known as Laplace's coefficient, is
the ratio of the heat capacity at constant pressure to the heat
capacity at constant volume. We can say that specific heat measures
the stiffness of the fluid. Substituting the value of $\bar{\mu}$ in
equation (\ref{51}), we get
\begin{eqnarray}\label{52}
\bar{P_r}&=&-\Gamma\left[P_{r0}\frac{b}{B_0}-2P_{r0}\left(\frac{\mu_0+P_{\bot0}}{\mu_0+P_{r0}}\right)\frac{c}{r}-\frac{P_{r0}}{\mu_0+P_{r0}}
D_1\right]T,\\\label{53}
\bar{P_\bot}&=&-\Gamma\left[P_{\bot0}\frac{b}{B_0}\left(\frac{\mu_0+P_{r0}}{\mu_0+P_{\bot0}}\right)+2\frac{c}{r}P_{\bot0}+\frac{P_{\bot0}}{\mu_0+
P_{\bot0}}D_1\right]T.
\end{eqnarray}
From equation (\ref{48}), we obtain
\begin{equation}\label{54}
\frac{\ddot{T}}{T}=\omega_\Sigma.
\end{equation}
Substituting the values of $\bar{\mu},~\bar{P_r}$ and $\bar{P_\bot}$ in Eq.(\ref{44}), we get the collapse equation as
\begin{eqnarray}\nonumber
0&=&\frac{1}{B_0}\left[\Gamma\left(-P_{r0}\frac{b}{B_0}-2P_{r0}\left(\frac{\mu_0+P_{\bot0}}{\mu_0+P_{r0}}\right)\frac{c}{r}
-\frac{P_{r0}}{\mu_0+P_{r0}}\right)'
+(\mu_0+P_{r0})\right.\\\nonumber&\times&\left.\left(\frac{a}{A_0}\right)'\right]T-\frac{A'_0}{B_0A_0}\left[(\mu_0+P_{r0})\frac{b}{B_0}
+2(\mu_0+P_{\bot0})\frac{c}{r}+D_1+\Gamma\left(P_{r0}
\frac{b}{B_0}\right.\right.\\\nonumber&+&\left.\left.2P_{r0}\left(\frac{\mu_0+P_{\bot0}}{\mu_0+P_{r0}}\right)\frac{c}{r}+\frac{P_{r0}}{\mu_0+P_{r0}}\right)\right]
+\frac{1}{B}\left[2(P_{r0}-P_{\bot0})\left(\frac{c}{r}\right)'\right.\\\nonumber&-&\left.2\Gamma\frac{1}{r}\left(P_{r0}\frac{b}{B_0}
+2P_{r0}\left(\frac{\mu_0+P_{\bot0}}{\mu_0+P_{r0}}\right)
\frac{c}{r}+\frac{P_{r0}}{\mu_0+P_{r0}}D_1\right)\right]T+2\Gamma\frac{T}{rB_0}\\\label{55}&&\left[P_{\bot0}
\frac{b}{B_0}\left(\frac{\mu_0+P_{r0}}{\mu_0+P_{\bot0}}\right)
+2P_{\bot0}\frac{c}{r}+\frac{P_{\bot0}}{\mu_0+P_{\bot0}}\right]+D_2\omega_\Sigma-D_3
\end{eqnarray}
This equation has a crucial significance in determining the stable/unstable regimes of the spherical star configuration.

\section{Stability Conditions Under Newtonian And Post-Newtonian Regimes}

In this section, we will find the stability restrains from the collapse equation
with the help of the adiabatic index at both N and pN regimes.

\subsection{Newtonian Limit}

At N approximation, we assume that geometry is defined with flat
background metric by which instability of the system is discussed in
the framework of $f(G)$ gravity. The metric coefficients are
constraint as $A_0=1,~B_0=1$. Also, we consider that energy density
is much greater than the pressure components. So we have,
$\mu_0>>P_{r0},~\mu_0>>P_{\bot0}$. Then Eq.(\ref{56}) takes the form
as
\begin{eqnarray}\nonumber
0&=&\Gamma\left[\left(P_{r0}\left(b+2\frac{c}{r}\right)\right)'+2\frac{P_{r0}}{r}\left(b+2\frac{c}{r}\right)-2\frac{P_{\bot0}}{r}\left(b+
   2\frac{c}{r}\right)\right]\\\label{56}&-&(\mu_0+P_{r0}){a'}-2(P_{r0}-P_{\bot0})\left(\frac{c}{r}\right)'-D_2\omega_\Sigma-D_3,
 \end{eqnarray}
which is the condition for the hydrostatic equilibrium. From this
equation, we can find a constraint on the adiabatic index for
unstable regions as
\begin{equation}\label{57}
\Gamma<\frac{(\mu_0+P_{r0})a'+2(P_{r0}-P_{\bot0})\left(\frac{c}{r}\right)'+D_2\omega_\Sigma+D_3}{\left(P_{r0}\left(b+2\frac{c}{r}\right)\right)'
+\frac{2}{r}\left(b+2\frac{c}{r}\right)(P_{r0}-P_{\bot0})},
\end{equation}
The system will be unstable if inequality (\ref{57}) holds, i.e., if
the effect of the numerator term is less than the term in
denominator. If the term in numerator is greater than the term in
denominator then the system will be in dynamical stability. We found
that $\Gamma$ plays an integral role in describing the stability of
the system.

\subsection{Post-Newtonian Limits}

For post-Newtonian background, we choose $A_0=1-m_0/r,~B_0=1+m_0/r$,
and relativistic corrections up to order $m_0/r$. For this
approximation, Eq.(\ref{55}) becomes
\begin{eqnarray}\nonumber
0&=&-\Gamma\left(1-\frac{m_0}{r}\right)\left[P'_{r0}\left(b-\frac{bm_0}{r}+2\frac{c}{r}\right)+P_{r0}\left(b'+\frac{2}{r^2}(c'r-
c)\right)\right]\\\nonumber&+&\mu_0a'\left(1-\frac{m_0}{r}\right)+\frac{\mu_0m_0}{r^2}(a'r-a)+P_{r0}a'-b\frac{m_0\mu_0}{r^2}
-2\mu_0m_0\frac{c}{r^3}
\\\nonumber&-&D_1\left(\frac{m_0}{r}+\frac{m_0^2}{r^3}\right)\left(1-\frac{m_0}{r}\right)+2\left(\frac{c}{r}\right)'P_{r0}
-2\left(\frac{c}{r}\right)'P_{\bot0}\left(1-\frac{m_0}{r}
\right)\\\label{58}&-&2\frac{\Gamma}{r}\left(b+2\frac{c}{r}\right)(P_{r0}-P_{\bot0})+D_2\omega_\Sigma+D_3.
\end{eqnarray}
From the above equation, we can extract the unstable constraint for our stellar interior as
\begin{eqnarray}\nonumber
\Gamma&<&\frac{\mu_0\left(a'-\frac{m_0a}{r^2}-m_0\frac{b}{r^2}+2c\frac{m_0}{r^3}\right)+P_{r0}a'+2\left(\frac{c}{r}\right)'(P_{r0}-
P_{\bot0})}{P'_{r0}L+P_{r0}M+\frac{2}{r}\left(b+2\frac{c}{r}\right)(P_{r0}-P_{\bot0})}\\\label{59}&+&\frac{D_2\omega_\Sigma+D_3-
D_1\left(\frac{m_0}{r}-\frac{m_0^2}{r^2}
+\frac{m_0^2}{r^3}\right)}{P'_{r0}L+P_{r0}M+\frac{2}{r}\left(b+2\frac{c}{r}\right)(P_{r0}-P_{\bot0})},
\end{eqnarray}
where $L$ and $M$ are defined as,
\begin{eqnarray}\nonumber
L&=&\left[b\left(1-2\frac{m_0}{r}\right)+2\frac{c}{r}\left(1-\frac{m_0}{r}\right)\right], \\\nonumber
M&=&\left[b'\left(1-\frac{m_0}{r}\right)+2\left(\frac{c}{r}\right)'+2\frac{m_0}{r}\left(\frac{c}{r}\right)'\right].
\end{eqnarray}
The anisotropic spherical system will be unstable until inequality
(\ref{59}) holds. The system will be in hydrostatic equilibrium when
both numerator and denominator terms have the same effects and the
system will enter the state of stability when the numerator term has
greater effect than the denominator term.

\section{Conclusion}

In this paper, we discussed a systematic analysis for a spherically
symmetric, locally anisotropic fluid which collapses adiabatically
and calculated the instability ranges for such fluid in the
background of $f(G)$ gravity. This objective is followed by
acquiring the field equations and the dynamical equation. In order
to witness the consequences of $f(G)$ gravity on the stability
ranges, we have considered a particular class of $f(G)$ family i.e.,
$f(G)=G+\alpha {G}^2$. Then the field equations and dynamical
equations are perturbed and after some calculations, we get the
collapse equation which helps to estimate limits of dynamical
instability. Bekenstein \cite{bekenstein1971hydrostatic} estimated
the gravitationally collapse of a charged fluid ball by generalizing
the Oppenheimer-Volkoff equations of hydrostatic equilibrium and
calculating a Christodoulou formula. Toyozawa and Shinozuka
\cite{toyozawa1980stability} analyzed the local and global
stabilities of an electron within adiabatic approximation. Ori
\cite{ori1997perturbative} studied the inner structure of a generic
rotating black hole by applying the small perturbation approach.
Moreover, stages of formation of the black hole and generalization
of the Price's analysis are also deliberated.

Blondin \emph{et al.} \cite{blondin2003stability} deliberated
stability of the spherical accretion shocks occurring in star
formation, core-collapse supernovae etc. Feng \emph{et al.}
\cite{yang2010collapse} investigated the aspects affecting stability
of the shallow tunnel face. Mostly, dynamical instability of compact
objects is described by the adiabatic index. We conclude that
dynamical instability given by $\Gamma$ depends upon anisotropic
pressure, energy density and the dark source terms of the $f(G)$
gravity. Moustakidis \cite{moustakidis2017stability} analyzed
stability criteria for the white dwarfs, neutron stars and
super-massive stars like compact objects in the similar fashion. We
concluded that
\begin{itemize}
  \item The system will be dynamically unstable until inequalities (\ref{57}) and (\ref{59}) hold.
  \item The compact objects will be in hydrostatic equilibrium when the inequalities (\ref{57}) and (\ref{59}) are violated.
\end{itemize}
It is noteworthy that if the dark source terms are eliminated then the constraints for dynamical instability in GR can be obtained.

\vspace{0.25in}

\noindent\textbf{Acknowledgement}

The work of MZB and ZY was supported by National Research Project
for Universities (NRPU), Higher Education Commission, Pakistan under
research project No. 8769/Punjab/ NRPU/R$\&$D/HEC/2017. The authors
would like to thank the anonymous reviewer for the valuable and
constructive comments and suggestions in order to improve the
quality of the paper.\\

\section{Appendix}

The extra curvature ingredients indicated by $\chi_i$ appearing in Eq.(\ref{14}) are given below
\begin{eqnarray}\nonumber
\chi_{1} &=&\frac{8}{AB^3R}\left(A\dot{B}R'-AB\dot{R'}+A'B\dot{R}\right),  \\\nonumber
\chi_2&=&\frac{4}{B^4R^2}(A^2R'^2-\dot{R}^2B^2-A^2B^2),  \\\nonumber
\chi_3&=&\frac{4}{AB^5R^2}(AB^2B'\dot{R^2}-3A^3B'R'^2+B'B^2A^3+2R''R'BA^3\\\nonumber&-&2AB^2\dot{B}\dot{R}R'-2ABRR'\dot{B^2}+2ARB^2\dot{B}\dot{R'}-
2A'B^2\dot{B}R\dot{R}) , \\\nonumber
\chi_4&=&\frac{4}{A^2B^3R^2}(3B^2\dot{B}\dot{R}^2-A^2R'^2\dot{B}+A^2B^2\dot{B}-2A^2B\dot{R}R''+2A^2B'R'\dot{R}
\\\nonumber&-&2ARA'\dot{B}R'+2ABRA'\dot{R'}-2A'^2BR\dot{R}).\\\nonumber
\end{eqnarray}
The values of $\phi_i$ arising in Eq.(\ref{16}) are given as
\begin{eqnarray}\nonumber
\phi_1&=&-\frac{8}{A^3BR}(A\dot{B}R'-AB\dot{R'}+A'B\dot{R}),  \\\nonumber
\phi_2&=&\frac{4}{A^6B^3R^2}(2AB^4\dot{A}\dot{B}R^2-2A^2B^4\ddot{B}R^2+2A^3B^3R^2A''-2A^3B^2R^2A'B'\\\nonumber&+&4RA^4B^3R''-\dot{R}^2A^2B^5
-4A^2B^4R\dot{R}\dot{B}+A^4B^3R'^2-4A^4B^2RR'B'\\\nonumber&-&A^4B^5),  \\\nonumber
\phi_3&=&\frac{4}{A^5B^2R^2}(-2A^2B^2R\dot{A}R''+\dot{A}B^4\dot{R}^2-A^2B^2\dot{A}R'^2-2AB^4\dot{R}\ddot{R}\\\nonumber&+&2A^2B^2A'R'\dot{R}-
A'B'RA^2B\dot{R}-2ARB^2A'^2\dot{R}+2A^2B^2RA'\dot{R'}\\\nonumber&-&2A^2BRA'\dot{B}R'),  \\\nonumber
\phi_4&=&\frac{4}{A^3B^2R^2}(A'B^2\dot{R}^2-3A^2A'R'^2+A^2B^2A'+2AB^2R'\ddot{R}\\\nonumber&-&2\dot{A}\dot{R}B^2R'+2AR\dot{B}^2R'-2ABR\dot{B}\dot{R'}
+2A'BR\dot{B}\dot{R}), \\\nonumber
\end{eqnarray}
The terms $\psi_i$ occurring in the Eq.(\ref{17}) are
\begin{eqnarray}\nonumber
\psi_1&=& \frac{4}{A^4B^3}(A^2BRR''-B^2R\dot{B}\dot{R}-A^2RR'B'), \\\nonumber
\psi_2&=& \frac{4}{A^3B^4}(AB^2R\ddot{R}-RB^2\dot{A}\dot{R}-RA^2A'R'), \\\nonumber
\psi_3&=&\frac{4}{A^5B^3}(A^2BRA'\dot{R'}-ABRA'^2\dot{R}-A^2BR\dot{A}R''+3RB^2\dot{A}\dot{B}\dot{R}\\\nonumber&+&A^2RB'R'\dot{A}-
AR\dot{B}B^2\ddot{R}-AB^2R\dot{R}\ddot{B}+A^2BR\dot{R}A''-A^2RA'B'\dot{R}),  \\\nonumber
\psi_4&=&\frac{4}{A^3B^5}(-ABRR'\dot{B^2}+AB^2R\dot{B}\dot{R'}-A^2BRA'R''+3RA^2A'B'R'\\\nonumber&-&ARB^2B'\ddot{R}+BRB'\dot{A}\dot{R}
+ARB^2R'\ddot{B}-A^2BRR'A''-B^2RR'\dot{A}\dot{B}),\\\nonumber
\end{eqnarray}
The expressions of $\eta_i$ which pop up in Eq.(\ref{15}) are given as
\begin{eqnarray}\nonumber
\eta_1&=&\frac{4}{A^4B^4R^2}(-A^2B^3\dot{B}\dot{R}R'+A^2B^4\dot{R}\dot{R'}-AB^3R^2A'\ddot{B}+A^2B^2R^2A'A''\\\nonumber&-&A^3B^2A'R'^2+B^3R^2A'
\dot{A}\dot{B}-A^2R^2BB'A'^2+A^3B^4A'),  \\\nonumber
\eta_2&=&\frac{4}{A^4B^4R^2}(-A^4B^2R'\dot{R'}+A^3B^2A'R'\dot{R}-A^2B^2R^2\dot{B}\ddot{B}+A^3BR^2\dot{B}A''\\\nonumber&+&A^2B^3\dot{B}\dot{R}^2+
AB^2R^2\dot{A}\dot{B}^2-A^3R^2\dot{B}B'A'+A^4B^3\dot{B}),  \\\nonumber
\eta_3&=& \frac{4}{A^3B^3R^2}(AB^2R^2\ddot{B}-AB^3\dot{R}^2+A^3BR'^2-B^2R^2\dot{A}\dot{B}+A^2R^2A'B'\\\nonumber&-&A^3B^3-A^2BR^2A''), \\\nonumber
\end{eqnarray}
The dark source terms $D_1,~D_2$ and $D_3$ in Eqs.(\ref{43}) and (\ref{44}) are
\begin{eqnarray}\nonumber
D_1 &=&-\frac{1}{A_0}\left(-2\frac{a}{A_0}+\frac{b}{B_0}+2\frac{c}{r}\right)\left[-8\alpha\frac{G''_0}{r^2B_0^4}(1-B_0^2)-8\alpha\frac{G'_0}{r^2B_0^5}
(-3B'_0\right.\\\nonumber&+&\left.B'_0B_0^2)\right]-\frac{1}{A_0}\left(\frac{b}{B_0}+2\frac{c}{r}\right)\frac{\alpha G_0^2}{2}+\alpha gG_0\frac{1}{A_0}-8\alpha\frac{G''_0}{A_0^2(r^2B_0^4)^2}[(2a\\\nonumber&+&2A_0^2c'-2aB_0^2-2bA_0B_0)r^2B_0^4-(A_0^2-   A_0^2B_0^2)(2rcB_0^4-4br^2B_0^3)]\\\nonumber&-&8\alpha\frac{g''}{r^2A_0B_0^4}(1-B_0^2)-
8\alpha\frac{G'_0}{A_0^3(r^2A_0B_0^5)^2}\left[r^2A_0B_0^5(
-6A_0^3B'_0c'-3b'A_0^3\right.\\\nonumber&-&\left.9aA_0^2B'_0+b'A_0^3B_0^2+2bA_0^3B_0B'_0+3aA_0^2B_0^2B'_0+2c''A_0^3B_0)-
(\right.\\\nonumber&-&\left.3A_0^3B'_0A_0^3B_0^2B'_0)(2rcA_0B_0^5+r^2aB_0^5+5br^2a0B_0^5) \right]-8\frac{\alpha}{A_0}\left(\frac{B'_0}{B_0^3}\right.\\\nonumber&-&\left.2\frac{1}{rB_0^2}\right)\left[\frac{g}{r^2A_0^4B_0^4}(r^2A_0^2B_0^2
A'_0A''_0-A'_0A_0^3B_0^2-r^2A_0^2B_0{A'_0}^2B'_0\right.\\\nonumber&+&\left.A_0^3B_0^4A'_0)-\frac{G'_0}{r^2A_0^4B_0^4}(-c'A_0^4B_0^2+cA_0^3B_0^2A'_0
+br^2A_0^3B_0{A''_0}+bA_0^4B_0^3\right.\\\nonumber&-&\left.br^2A_0^3A'_0B'_0)+\frac{g'}{r^2A_0^3B_0^3}(A_0^3B_0+r^2A_0^2A'_0B'_0-A_0^3B_0^3
-r^2A_0^2B_0A''_0)\right]\\\nonumber&-&8\frac{\alpha}{A_0B_0^2}\left[ \frac{g}{(r^2A_0^4B_0^4)^2}[(r^2A_0^4B_0^4)(r^2A_0^2B_0^2 A'_0 A''_0+
2r^2A_0^2B_0A'_0B'_0A''_0\right.\\\nonumber&+&\left.2rA'_0A_0^2B_0^2A''_0+2r^2A_0B_0^2{A'_0}^2A''_0+r^2A_0^2B_0^2{A''_0}^2
-2A_0^3B_0^2A'_0\right.\\\nonumber&-&\left.2A_0^3A'_0B_0B'_0-3A_0^2B_0^2{A'_0}^2
-A_0^3B_0^2A''_0-2r^2A_0^2B_0{A'_0}^2B'_0\right.\\\nonumber&-&\left.r^2A_0^2B_0{A'_0}^2B''_0-2rA_0^2B_0{A'_0}^2B'_0-2r^2A_0B_0{A'_0}^3B'_0+r^2A_0^2
{A'_0}^2{B'_0}^2\right.\\\nonumber&+&\left.4A_0^4B_0^3B'_0
+3A_0^2B_0^4{A'_0}^2+A_0^3B_0^4A''_0)+(2rA_0^4B_0^4+4r^2A_0^3B_0^4B'_0\right.\\\nonumber&+&\left.4r^2A_0^4B_0^3B'_0)(r^2A_0^2B_0^2A'_0A''_0-A_0^3A'_0B_0^2-r^2A_0^2B_0{A'_0}^2B'_0
+A_0^3B_0^4A'_0)]\right.\\\nonumber&+&\left.\frac{g'}{r^2A_0^4B_0^4}(r^2A_0^2B_0^2A'_0A''_0-A_0^3B_0^2A'_0-r^2A_0^2B_0{A'_0}^2B'_0+A_0^3B_0^4A'_0)
\right.\\\nonumber&+&\left.\frac{G'_0}{(r^2A_0^4B_0^4)^2}[r^2A_0^4B_0^4(-c'B_0^2A_0^4+cA'_0A_0^3B_0^2+r^2bA_0^3B_0A''_0-r^2bA_0^3A'_0B'_0
\right.\\\nonumber&+&\left.bA_0^4B_0^3)'-(r^2A_0^4B_0^4)'(-c'B_0^2A_0^4+cA'_0A_0^3B_0^2+r^2bA_0^3B_0A''_0+bA_0^4B_0^3\right.\\\nonumber&
-&\left.r^2bA_0^3A'_0B'_0)]
+\frac{G''_0}{r^2A_0^4B_0^4}(-c'B_0^2A_0^4+cA'_0A_0^3B_0^2+r^2bA_0^3B_0A''_0\right.\\\nonumber&-&\left.r^2bA_0^3A'_0B'_0+bA_0^4B_0^3)
+\frac{g'}{(r^2A_0^3B_0^3)^2}[r^2A_0^3B_0^3(A_0^3B_0+r^2A_0^2A'_0B'_0\right.\\\nonumber&-&\left.A_0^3B_0^3-r^2A_0^2B_0A''_0)'-
(r^2A_0^3B_0^3)'(A_0^3B_0
+r^2A_0^2A'_0B'_0-A_0^3B_0^3\right.\\\nonumber&-&\left.r^2A_0^2B_0A''_0)]+\frac{g''}{r^2A_0^3B_0^3}(A_0^3B_0
+r^2A_0^2A'_0B'_0-A_0^3B_0^3-r^2A_0^2B_0A''_0)
\right.\\\nonumber&+&\left.\frac{b}{A_0B_0^3}\left(\frac{G'_0}{r^2A_0B_0^2}(-3A'_0+A'_0B_0^2)\right)+16\alpha\frac{c}{r^2A_0}(-G''_0\frac{A'_0}
{A_0B_0^4}\right.\\\nonumber&+&\left.\frac{G'_0}{A_0B_0^5}(3A'_0B'_0-A''_0B_0))\right], \\\nonumber
D_2&=&8\frac{\alpha}{r^2A_0^2B_0}\left[\frac{g}{A_0^4B_0^4}(r^2A_0^2B_0^2A'_0A''_0-A_0^3B_0^2A'_0-r^2A_0^2B_0B'_0{A'_0}^2\right.\\\nonumber&+&\left.
A_0^3A'_0B_0^4)+\frac{G'_0}{A_0^4B_0^4}(-c'B_0^2A_0^4+cA'_0A_0^3B_0+r^2bA_0^3B_0A''_0-r^2bA_0^3A'_0B'_0\right.\\\nonumber&+&\left.bA_0^4B_0^3)
+\frac{g'}{A_0^3B_0^3}(A_0^3B_0+ r^2A_0^2A'_0B'_0-A_0^3B_0^3-r^2A_0^2B_0A''_0)\right]\\\nonumber&-&8\alpha\frac{g}{r^2A_0^3B_0^4}\left(
-2\frac{B'_0}{B_0}+\frac{A'_0}{A_0}
+\frac{2}{r}\right)(2r^2A''_0B_0-2r^2A'_0B'_0+A_0B_0-\\\nonumber&-&4rA_0B'_0-A_0B_0^3)-16\alpha c\frac{G'_0}{r^2A_0^2B_0^3}(
-2\frac{B'_0}{B_0}+\frac{A'_0}{A_0}+\frac{2}{r})\\\nonumber&-&8\alpha\frac{g}{(r^2A_0^5B_0^2)^2B_0^3}[(r^2A_0^5B_0^2)(2r^2A_0^3B_0^3A''_0
-2r^2A_0^3B_0^2A'_0B'_0+A_0^4B_0^3\\\nonumber&-&4rA_0^4B_0^2B'_0-A_0^4B_0^5)'-(r^2A_0^5B_0^2)'(2r^2A_0^3B_0^3A''_0-
2r^2A_0^3B_0^2A'_0B'_0\\\nonumber&+&A_0^4B_0^3-4rA_0^4B_0^2B'_0-A_0^4B_0^5)]-8\alpha\frac{g'}{r^2A_0^5B_0^5}(2r^2A_0^3B_0^3A''_0
\\\nonumber&-&2r^2A_0^3B_0^2A'_0B'_0
+A_0^4B_0^3-4rA_0^4B_0^2B'_0-A_0^4B_0^5)-8\alpha\frac{G'_0}{(r^2A_0^3B_0^2)^2B_0^3}
\\\nonumber&\times&[(r^2A_0^3B_0^2)(2{c'}A_0B_0^2+4cA_0B_0{B'_0}+2cB_0^2A'_0)
-2cA_0B_0^2]-8\alpha\frac{G''_0}{r^2A_0^3B_0^5}\\\nonumber&\times&(2cA_0B_0^2)-16\alpha[g\frac{B'_0}{r^2A_0^2B_0^4}-c\frac{G''_0}{r^2A_0^2B_0^3}-\frac{G'_0}
{r^2A_0^2B_0^4}(-cB'_0+b)],\\\nonumber
D_3 &=& -2\frac{B'_0}{B_0^4}\left[-\alpha G_0^2\frac{B_0^2}{2}\left(\frac{b'}{B'_0}\right)-\alpha gG_0B_0^2-8\alpha\frac{G'_0}{r^2A_0B_0^2}\left(-3A'_0\left(2r(\frac{c}{r})'\right.\right.\right.\\\nonumber&+&\left.\left.\left.\frac{a'}{A'_0}-\frac{a}{A_0}
+\frac{b'}{B'_0}\right)+A'_0B_0^2\left(
-\frac{a}{A_0}-\frac{a'}{A'_0}-2\frac{c}{r}+\frac{b'}{B'_0}\right)\right)\right]
+\frac{A'_0}{A_0B_0^3}\left[\right.\\\nonumber&-&\left.\alpha G_0^2\frac{B_0^2}{2}\left(\frac{a'}{A'_0}-\frac{a}{A_0}\right)-\alpha gG_0B_0^2-8\alpha\frac{G'_0}{r^2A_0B_0^2}\left(
-3A_0\left(2c'+2\frac{a'}{A'_0}\right.\right.\right.\\\nonumber&-&\left.\left.\left.2\frac{a}{A_0}-2\frac{c}{r}\right)+A'_0B_0^2\left(-2\frac{a}{A_0}
-2\frac{c}{r}\right)\right)-8\alpha\frac{g'}{r^2A_0B_0^2}(-3A'_0+A'_0B_0^2)\right]\\\nonumber&+&\frac{2}{rB_0^3}\left[
-\alpha G_0^2\frac{B_0^2}{2}\left(c'-\frac{c}{r}\right)-\alpha gG_0B_0^2-8\alpha\frac{G'_0}{r^2A_0B_0^2}\left(-3A'_0\left(3c'+\frac{a'}{A'_0}
\right.\right.\right.\\\nonumber&-&\left.\left.\left.
\frac{a}{A_0}-3\frac{c}{r}\right)+A'_0B_0^2\left(-\frac{a}{A_0}-\frac{a'}{A'_0}-3\frac{c}{r}+c'\right)\right)-8\alpha\frac{g'}{r^2A_0B_0^2}
(-3A'_0\right.\\\nonumber&+&\left.A'_0B_0^2)\right]+\frac{A'_0}{A_0^3B_0}\left[\alpha G_0^2\frac{A_0^2}{2}\left(-\frac{a}{A_0}+\frac{a'}{A'_0}\right)+\alpha gG_0A_0^2-
8\alpha\frac{G''_0}{r^2B_0^4}\left(A_0^2\left(\right.\right.\right.\\\nonumber&-&\left.\left.\left.\frac{a}{A_0}+2c'-2\frac{c}{r}
+\frac{a'}{A'_0}\right)-A_0^2B_0^2\left(-\frac{a}{A_0}-2\frac{c}{r}
+\frac{a'}{A'_0}\right)\right)-8\alpha \frac{g''}{r^2B_0^4}(A_0^2\right.\\\nonumber&-&\left.A_0^2B_0^2)-8\alpha\frac{G'_0}{r^2B_0^5}\left[-3A_0^2B'_0\left(2c'+\frac{b'}{B'_0}
-2\frac{c}{r}-\frac{a}{A_0}+\frac{a'}{A'_0}\right)\right.\right.\\\nonumber&+&\left.\left.A_0^2B_0^2B'_0\left(\frac{b'}{B'_0}-\frac{a}{A_0}-
2\frac{c}{r}+\frac{a'}{A'_0}\right)
+2c''A_0^2B_0\right]-8\alpha\frac{g'}{r^2B_0^5}(-3A_0^2B_0\right.\\\nonumber&+&\left.A_0^2B_0^2B'_0)\right]+\frac{1}{B_0^3}\left[-\alpha B_0B'_0G_0^2\frac{b'}{B'_0}-2\alpha gG_0B_0B'_0-\alpha B_0^2(gG_0)'\right.\\\nonumber&-&\left.8\alpha\frac{G'_0}{(r^2A_0^3B_0^2)^2B_0^3}\left[r^2A_0^3B_0^2\left(-6c''A_0^2A'_0
-3A''_0A_0^2\left(2c'+\frac{a''}{A''_0}-\frac{a}{A_0}\right.\right.\right.\right.\\\nonumber&-&\left.\left.\left.2\frac{c}{r}\right)
-6A_0{A'_0}^2\left(2c'+2\frac{a'}{A'_0}-2\frac{a}{A_0}-2\frac{c}{r}\right)
+2A_0^2A'_0B_0B'_0\left(\frac{b'}{B'_0}+\frac{a'}{A'_0}\right.\right.\right.\\\nonumber&-&\left.\left.\left.\frac{a}{A_0}-2\frac{c}{r}\right)
+2A_0{A'_0}^2B_0^2\left(-2\frac{a}{A_0}+2\frac{a'}{A'_0}
-2\frac{c}{r}\right)+A''_0A_0^2B_0^2\left(-\frac{a}{A_0}\right.\right.\right.\\\nonumber&+&\left.\left.\left.\frac{a''}{A''_0}
-2\frac{c}{r}\right)\right)-\left(2rA_0^3B_0^2\left(-3\frac{c}{r}+c'
-3\frac{a}{A_0}\right)+2r^2A_0^3B_0B'_0\left(-\frac{c}{r}\right.\right.\right.\\\nonumber&+&\left.\left.\left.\frac{b'}{B'_0}-
3\frac{a}{A_0}\right)+3r^2A_0^2A'_0B_0^2\left(-2\frac{c}{r}-4\frac{a}{A_0}
+\frac{a'}{A'_0}\right)\right)\right]\\\nonumber&-&8\alpha\frac{g'}{(r^2A_0^3B_0^2)^2B_0^3}\left[(r^2A_0^3B_0^2)(-3A_0^2A'_0+A_0^2B_0^2A'_0)'-
(r^2A_0^3B_0^2)'(\right.\\\nonumber&-&\left.3A_0^2A'_0+A_0^2B_0^2A'_0)\right]-8\alpha\frac{G''_0}{r^2A_0^3B_0^5}\left[-3A_0^2A'_0
\left(c'+\frac{a'}{A'_0}-\frac{a}{A_0}
-2\frac{c}{r}\right)\right.\\\nonumber&+&\left.A_0^2A'_0B_0^2\left(-\frac{a}{A_0}+\frac{a'}{A'_0}-2\frac{c}{r}\right)\right]
-8\alpha\frac{g''}{r^2A_0^3B_0^5}
(-3A'_0A_0^2+A_0^2A'_0B_0^2)\\\nonumber&+&\frac{2}{r^3B_0}\left[r^2\alpha\frac{G_0^2}{2}\left(-\frac{c}{r}+c'\right)+r^2\alpha gG_0-8\alpha g\frac{rB'_0}{A_0^2B_0^3}-8r\alpha A'_0\frac{G''_0}{A_0B_0^4}\left(2c'\right.\right.\\\nonumber&+&\left.\left.3\frac{a'}{A'_0}-2\frac{c}{r}-\frac{a}{A_0}\right)-8\alpha r\frac{g''A'_0}
{A_0B_0^4}+8\alpha\frac{G'_0}{A_0^3B_0^5}\left(-rc''A_0^2A'_0B_0\right.\right.\\\nonumber&+&\left.\left.3rA_0^2A'_0B'_0\left(
2c'+\frac{a'}{A'_0}+\frac{b'}{B'_0}-2\frac{c}{r}-
\frac{a}{A_0}\right)-rA_0^2B_0A''_0\left(\frac{a''}{A''_0}-2\frac{c}{r}\right.\right.\right.\\\nonumber&-&\left.\left.\left.\frac{a}{A_0}+2c'\right)\right)+ 8\alpha\frac{g'}{A_0B_0^5}(3rA'_0B'_0-rA''_0B_0)\right]+\frac{b}{B_0}\left[
\frac{1}{2}\alpha G_0^2\frac{A'_0}{A_0B_0}\right.\\\nonumber&+&\left.\alpha\frac{G_0^2}{rB_0}\right]
+\alpha\frac{b}{B_0}\left[240 {G'_0}\frac{A'_0B'_0}{r^2A_0B_0^6}-120 {G'_0}\frac{{A'_0}^2}{r^2A_0^2B_0^5}-240{G'_0}\frac{A'_0}{r^2A_0B_0^5}
\right.\\\nonumber&-&\left.48 {G'_0}\frac{A'_0B'_0}{r^2B_0^4}+24 {G'_0}\frac{{A_0^2}'}{r^2A_0B_0^3}+48 {G'_0}\frac{A'_0}{r^3B_0^3}+32{G''_0}\frac{A'_0}{r^2A_0B_0^5}\right.\\\nonumber&-&\left.16
{G''_0}\frac{A'_0}{r^2A_0B_0^3}-120 {G'_0}\frac{A'_0B'_0}{r^2A_0B_0^6}+24 {G'_0}\frac{A'_0B'_0}{r^2A_0B_0^4}-96G'_0
\frac{A''_0}{r^2A_0B_0^5}\right.\\\nonumber&-&\left.192G'_0\frac{{A'_0}^2}{r^2A_0^2B_0^5}+48G'_0\frac{A'_0B'_0}{r^2A_0B_0^4}
+32G'_0\frac{{A'_0}^2}{r^2A_0^2B_0^3}
+16G'_0\frac{A''_0}{r^2A_0B_0^3}\right.\\\nonumber&+&\left.192G'_0\frac{A'_0}{r^3A_0B_0^5}+240G'_0\frac{A'_0B'_0}{r^2A_0B_0^6}
+288G'_0\frac{{A_0}'^2}{r^2A_0^2B_0^5}
-32G'_0\frac{A'_0}{r^3A_0B_0^3}\right.\\\nonumber&-&\left.48G'_0\frac{A'_0B'_0}{r^2A_0B_0^4}-48G'_0\frac{{A'_0}^2}{r^2A_0^2B_0^3}-
96G''_0\frac{A'_0}{rA_0B_0^2}
+16G''_0\frac{A'_0}{r^2A_0B_0}\right.\\\nonumber&+&\left.64G''_0\frac{A'_0}{r^2A_0B_0^5}-240G'_0\frac{A'_0B'_0}{r^2A_0B_0^6}
+64G'_0\frac{A''_0}{r^2A_0B_0^5}\right].
\end{eqnarray}

\end{document}